\documentclass[twocolumn,showpacs,preprintnumbers,amsmath,amssymb,prl]{revtex4}
\usepackage{graphicx}% Include figure files
\usepackage{dcolumn}% Align table columns on decimal point
\usepackage{bm}% bold math
\usepackage[]{epsfig}

\begin{document}
\title{Collective rearrangement at the onset of flow of a polycrystalline
hexagonal columnar phase}
\author{Teresa Bauer, Julian Oberdisse and Laurence Ramos$^*$}
\affiliation{Laboratoire des Collo\"{\i}des, Verres et
Nanomat\'{e}riaux (UMR CNRS-UM2 5587), CC26, Universit\'{e}
Montpellier 2, 34095 Montpellier Cedex 5, France}

\email{ramos@lcvn.univ-montp2.fr}
\date{\today}

\begin{abstract}

Creep experiments on polycrystalline surfactant hexagonal columnar
phases show a power law regime, followed by a drastic fluidization
before reaching a final stationary flow. The scaling of the
fluidization time with the shear modulus of the sample and stress
applied suggests that the onset of flow involves a bulk
reorganization of the material. This is confirmed by X-ray
scattering under stress coupled to \textit{in situ} rheology
experiments, which show a collective reorientation of all
crystallites at the onset of flow. The analogy with the fracture
of heterogeneous materials is discussed.

\end{abstract}

\maketitle

% body of paper here

Yield stress fluids are materials that respond elastically below a
yield stress, $\sigma_y$, and flow as liquids for stresses above
$\sigma_y$ \cite{BarnesJNNFM1999}.  Typical materials that fall in
this category include granular media, dry foams or dense colloidal
suspensions and emulsions. Very recently, B\'{e}cu \textit{et al.}
\cite{BecuPRL2006} have shown that the behavior of concentrated
emulsions at the onset of flow depends crucially on the
microscopic interactions between the droplets, thus underlining
the non universal nature of the solid to fluid transition of yield
stress fluids. Several fundamental issues are still unclear
regarding this transition. These include in particular the
possible prediction of if and when a material will flow, from its
behavior prior to flow. In addition, whether the solid to fluid
transition is accompanied by structural modifications of the yield
stress fluids remains largely unknown.

We perform creep experiments in order to investigate the onset of
flow of  soft polycrystalline hexagonal columnar phases. Upon
application of a small and constant stress, $\sigma$, the material
deforms plastically in a similar manner as a solid \cite{Modulus}.
By contrast, under high stress, it behaves as a fluid and flows
with a constant shear rate that depends on $\sigma$
\cite{ShearMelting1,ShearMelting2}. Quite remarkably, for an
intermediate value of $\sigma$, an intriguing regime is observed
during which the material, after an incubation time that depends
on both $\sigma$ and the sample shear modulus, evolves suddenly
from a power law creep to a flow regime, through a dramatic
decrease of the viscosity. Thanks to a combination of Synchrotron
X-ray scattering experiments under stress and \textit{in situ}
rheology measurements, we are able to correlate the solid to fluid
transition with structural modifications of the material, thus
shedding light on the mechanisms involved at the onset of flow of
some complex fluids.

\begin{figure}
\includegraphics{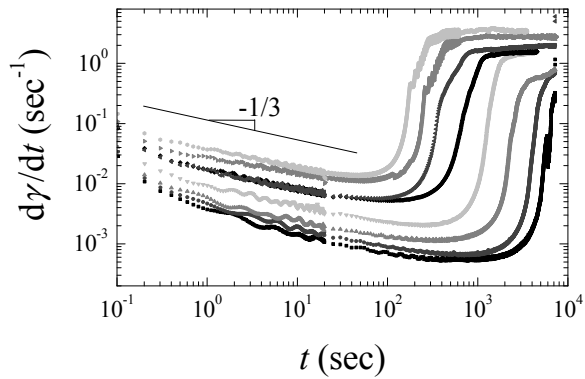}% Here is how to import EPS art
\caption{Time-evolution of the shear-rate in creep experiments for
a sample with an elastic shear modulus $G_0=2240\, \rm{Pa}$. At
time $0$ a constant stress $\sigma = 12, 13, 15, 19, 21, 22, 30,
35 \, \rm{Pa}$ (from bottom to top) is applied to the sample.}
\label{FIG:1}
\end{figure}

The experimental system is a lyotropic hexagonal phase consisting
of infinitely long oil tubes which are stabilized by a surfactant
monolayer and immersed in water. The tubes of uniform radius
arrange on a triangular array and form, therefore, a
two-dimensional columnar crystal. The samples' composition and
elasticity have been described elsewhere \cite{Swelling, Modulus}.
Four samples with variable shear moduli, $G_0$, ranging from $250$
to $3200 \, \rm{Pa}$, have been investigated. Rheology experiments
are performed on a stress-controlled Paar Physica UDS 200
rheometer in a Couette geometry with a gap of $1$ mm. Once loaded
in the cell, the sample is systematically cooled below the
hexagonal to isotropic fluid  phase transition temperature ($\sim
10 ^{\circ} \rm{C}$), and then warmed up to room temperature. A
polycrystal, with an isotropic distribution for the crystallites
orientation, is thus rapidly recovered. At time $t=0$, we then
apply a constant shear stress, $\sigma$, to the material and
record the time evolution of the strain, $\gamma$. An
instantaneous jump of $\gamma$ is measured, revealing the elastic
nature of the material at high frequency, which is followed by a
monotonic increase of $\gamma$. The evolution of the instantaneous
shear rate, $\dot{\gamma}$, calculated as the time derivative of
$\gamma$, is shown in fig.\ \ref{FIG:1} for several stresses, and
displays three distinct regimes. Initially, $\dot{\gamma}$
decreases as a power law with time, indicating a continuous
increase of the instantaneous viscosity, $\eta$, defined as
$\sigma / \dot{\gamma}$. This power law regime is followed by a
very abrupt increase of $\dot{\gamma}$ of more than $2$ orders of
magnitude, until the shear rate reaches a constant and high value,
which corresponds to a stationary flow with a lower viscosity. For
$t$ larger than $0.1 \, \rm{sec}$ (the time resolution of our
experiments), and over at least three orders of magnitude, we find
in the  power law regime $\dot{\gamma} \sim t^p$ with $p=-0.34 \pm
0.05$, whatever $\sigma$ and $G_0$. This regime recalls the
primary creep regime, the so-called Andrade creep \cite{Andrade}
measured in many crystalline and non-crystalline materials.
However, the catastrophic acceleration of the shear rate that
follows the primary creep, which illustrates a drastic decrease of
the viscosity, is neither observed for metallic materials, nor for
yield stress fluids such as foams \cite{CohenAddadPRL2004} or
colloidal glasses \cite{PetekidisJPCM2004}, where the transition
from creep to flow is smooth. The features shown in fig.\
\ref{FIG:1} appears rather unique for a complex fluid, but in fact
display intriguing analogies with the fracture under constant load
of many composite heterogeneous materials \cite{GuarinoEPJB2002,
NechadPRL2005}. We define the fluidization time, $\tau_f$, as the
time at which $\dot{\gamma}$ is minimum (i.e. the instantaneous
viscosity is maximum).

\begin{figure}
\includegraphics{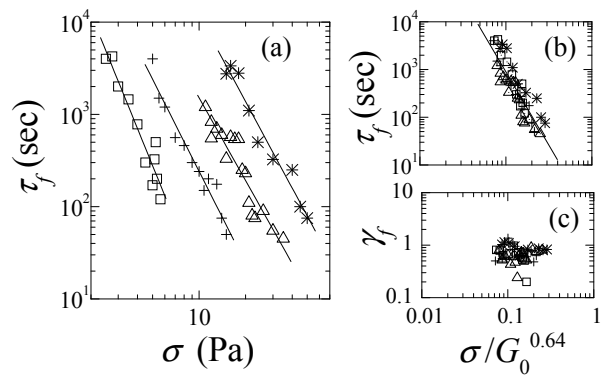}% Here is how to import EPS art
\caption{Fluidization time, $\tau_f$, as a function of (a) the
stress applied $\sigma$, and (b)  $\sigma / G_0^{0.64}$ (symbols),
for samples with different shear moduli, $G_0 = 250$ Pa (squares),
$760$ Pa (crosses), $2240$ Pa (triangles) and $3200$ Pa (stars) ;
the line in (b) is a power law fit yielding an exponent $-3.1 \pm
0.2$. (c) Strain at $t=\tau_f$ as a function of $\sigma /
G_0^{0.64} $ (same symbols as in (a,b)).} \label{FIG:2}
\end{figure}

We report the stress dependence of the fluidization time in fig.\
\ref{FIG:2}a, for several samples with different elasticities. We
find that, for all samples, $\tau_f$ decreases as $\sigma$
increases, and that curves for softer samples are shifted to lower
stresses. Interestingly, all data points can be put onto a single
master curve if the stress is renormalized by a power law of the
shear modulus, $G_0^m$. We find the best collapse of the data for
$m=0.64$. For this value, the master curve can be reasonably well
fitted with a power law $\tau_{f} \sim (\sigma /
G_0^{0.64})^{(-3.1\pm 0.2)}$ (fig.\ \ref{FIG:2}b). This scaling
implies that there is no divergence of $\tau_f$ with stress. It
 indicates that the material does not show any detectable
yield stress in our experimental time window, in contrast with the
findings for micellar polycrystals for instance
\cite{EiserEPJE2000}. On the other hand, we have shown previously
that the shear modulus of the polycrystal scales as $R^{-3}$,
where $R$ is the radius of the surfactant tubes \cite{Modulus}.
Hence, the normalized stress $\sigma / G_0^{0.64}$ scales roughly
as $\sigma \times R^2$. We expect the number density of tubes in
the material to scale as $1/R^2$. Thus, the master curve indicates
that $\tau_f$ depends only on the stress applied per tube.
Interestingly, despite variation of $\tau_f$ of two orders of
magnitude, the strain $\gamma_f$ at time $t=\tau_f$ is
approximately constant. We find: $\gamma_f \simeq (69 \pm 23) \%$,
irrespective of the stress applied and of the sample elasticity
(fig.\ \ref{FIG:2}c). This shows that the onset of flow results
from a strain controlled mechanism. Note that a critical strain
similar to ours (i.e. of the order of $1$) is typical for
shear-induced order in hard sphere suspensions
\cite{AckersonPRL1988} or to completely fluidize a foam sample.
Both a critical strain of $1$ and the remarkable dependence of the
stress with the number of tube suggests that the fluidization time
reflects a bulk mechanism that involves the whole sample and not
an interfacial property. This is supported by the results
described below.

In order to gain a better understanding on the mechanisms at work
in the creep regime and at the onset of flow, we have coupled
time-resolved Synchrotron small-angle X ray scattering (SAXS)
under stress to \textit{in situ} rheological measurements.
Experiments have been performed on the ID-2 beamline at the ESRF,
Grenoble, France. A stress-controlled Haake RS300 rheometer
equipped with a Couette cell is used. The scattering profiles are
recorded for two geometries, the incident beam being either radial
or tangential with respect to the cell. Only the softer sample
($G_0= 250\, \rm{Pa}$) has been investigated. A procedure
analogous to the one used for the rheology experiments has been
performed in order to ensure a reproducible polycrystalline
initial state corresponding to an isotropic distribution for the
orientation of the crystallites, as confirmed by isotropic rings
found in the SAXS patterns for both radial and tangential
geometries. We can evaluate that the size of the crystallites is
of the order of $1\, \mu\rm{m}$ \cite{NoteCrystallites}. Taking
into account the size of the incident beam ($100\, \mu\rm{m}$) and
that of the gap ($1$ mm), we conclude that the SAXS spectra
reflects the orientation distribution of a large assembly of
crystallites.

\begin{figure}
\includegraphics{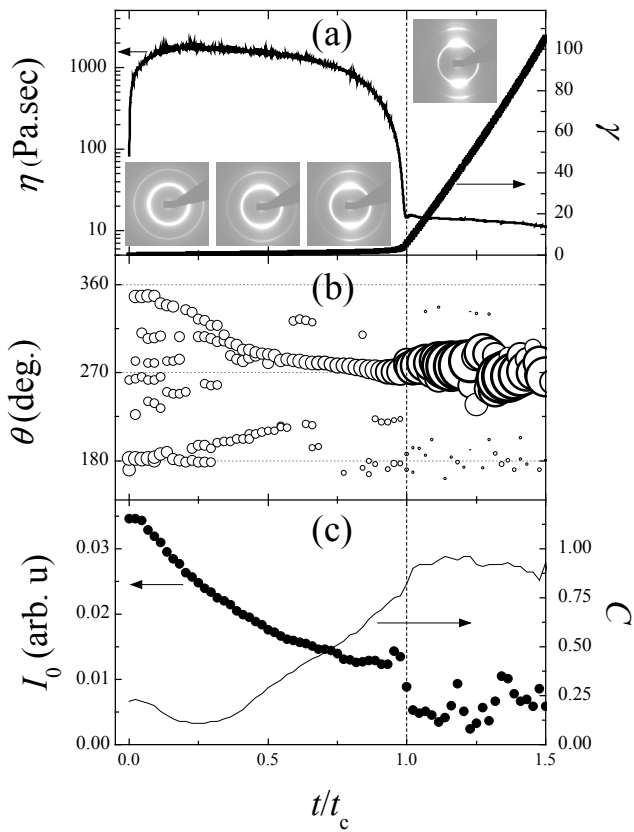}% Here is how to import EPS art
\caption{Combined rheological and structural measurements in a
creep experiment. The sample elasticity is $G_0=250\, \rm{Pa}$ and
the stress is $\sigma=2\, \rm{Pa}$.  The time is normalized by the
time $t_c$ at which the sample flows with a constant shear rate.
The dashed vertical line marks the beginning of a stationary flow
regime. (a) \textit{In situ} rheological measurements, strain
(thick line) and instantaneous viscosity (thin line), and SAXS
patterns taken from left to right at $t/t_c=0.25, 0.75, 0.98$ and
$1.14$. The grey scale is linear and identical for all images. (b)
Time evolution of the angular position of the local intensity
maxima in an azimuthal scan. The bubble size scales with
intensity. (c) Time evolution of the contrast $C$ (line) and the
scattered intensity at $\theta = (180 \pm 2)$ deg (solid circles).
} \label{FIG:3}
\end{figure}

A typical series of radial scattering patterns are shown in fig.\
\ref{FIG:3}a together with the strain, $\gamma$, and the
instantaneous viscosity, $\eta$. The time evolution of $\eta$
clearly shows the three successive regimes (power law creep,
fluidization and stationary flow) described previously. In  fig.\
\ref{FIG:3}a, the time is normalized by $t_c$, the time at which
the material starts to flow with a stationary shear rate. The SAXS
patterns exhibit isotropic rings at the beginning of the
experiment, and then converts into a markedly anisotropic signal
in the flow regime, with higher intensity along the vorticity
direction ($\theta=90$ and $270$ deg). To quantify the time
evolution of the patterns and correlate it with that of the
rheological properties, we compute the angular dependence of the
scattered intensity integrated around the first Bragg peak
(azimuthal scan), and identify the angular position of the local
maxima. Figure\ \ref{FIG:3}b displays the evolution of the
position and intensity of these maxima, for $\theta$ in the range
$160-340$ deg ($\theta = 0$ and $180$ deg. corresponds to the
velocity direction). The experimental data clearly show that, as
soon as the systems flows ($t/t_c \geq 1$), the surfactant tubes
are in majority aligned along the flow (maxima around $270$ deg),
although the angular distribution for the tubes orientation is
wide ($\Delta \theta \sim 35$ deg). We note that the angular
width, $\Delta \theta$, and the shear rate measured in the flow
regime ($ \dot{\gamma} = 0.14 \, \rm{sec}^{-1}$ for $\sigma=2$ Pa)
are in quantitative agreement with our previous findings
\cite{ShearMelting1,ShearMelting2} concerning both the continuous
decrease of $\Delta \theta$ as $ \dot{\gamma}$ increases and the
stress dependence of $ \dot{\gamma}$, suggesting that the flow
regime investigated here belongs to the shear-thinning regime
previously determined for $ \dot{\gamma}$ in the range $(1-40) \,
\rm{sec}^{-1}$. We calculate the contrast, $C$, defined as
$(I_{max} - I_{min})/(I_{max} + I_{min})$, where $I_{max}$ and
$I_{min}$ are respectively the maximal and minimal intensity along
the azimuthal scan profile ; $C$ varies between $0$, for a
perfectly isotropic distribution of the crystallite orientation,
and $1$ for a totally anisotropic distribution. As shown in fig.\
\ref{FIG:3}c, $C$ is small ($\simeq 0.25$) at the beginning of the
experiment, and after a slight decrease \cite{NoteContrast},
increases continuously, until it reaches a constant and high
value, $C \simeq 0.92$, at $t=t_c$.

Because each experiment probes a given ensemble of crystallites,
the exact patterns of the peak position along a scan change from
one experiment to another one ; in particular the intensity around
$\theta=90$ and $270$ deg does not exhibit a clear trend in the
power law creep regime. Two experimental observations are
nevertheless robust. The first one is the systematic concomitance
of the contrast $C$ reaching its plateau value with the onset of
flow. This findings demonstrate that the onset of flow is
characterized by a collective rearrangement of all crystallites
such that they all orient along the flow direction, hence the flow
is homogeneous, rather than by wall slippage (as observed in the
vicinity of the yield stress for a microgel paste
\cite{MeekerPRL2004}), or shear localization (as observed in
adhesive emulsions \cite{BecuPRL2006}, or wet granular media
\cite{HuangPRL2005}). This mechanism is moreover in agreement with
the rheology experiments (fig.\ \ref{FIG:2}) which indicates a
bulk mechanism. Note in addition that data recorded in tangential
geometry do not show any significant variation with time:
isotropic rings are always measured, showing that, in the
direction perpendicular to the flow, the sample texture remains
powder-like, and the width of the Bragg peaks do not vary,
indicating a fixed average crystallites size. These observations
are consistent with the physical picture that only the orientation
of crystallites changes in the creep and at the onset of flow.
This also compares favorably with simulations of polycrystals that
show rotation of individual grains under a constant external load
\cite{AhluwaliaPRL2003}.

\begin{figure}
\includegraphics{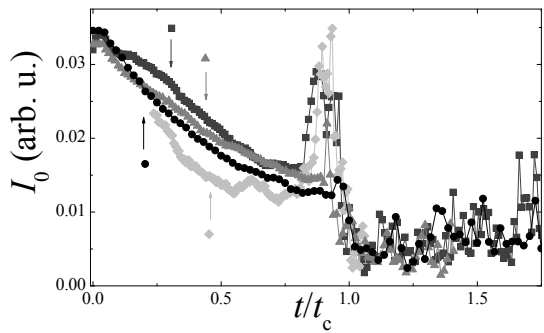}% Here is how to import EPS art
\caption{Scattered intensity in the direction perpendicular to the
flow ($\theta = (180 \pm 2) \, \rm{deg}$) for four distinct
experiments, with applied stress of $2$ (circles and up
triangles), $2.5$ (squares) and $3 \, \rm{Pa}$ (diamonds). The
time is normalized by the time $t_c$ at which the sample flows,
and the intensity by the scattered intensity integrated over all
angles in the flow regime. The arrows indicate the normalized
fluidization time, $\tau_f / t_c$.} \label{FIG:4}
\end{figure}

The second robust experimental observation concerns $I_0$, the
scattered intensity for $\theta = (180 \pm 2)$ deg. $I_0$ arises
from the scattering of tubes that are perpendicular to the flow
(tubes parallel to both the vorticity direction and the velocity
gradient direction contribute to $I_0$), and resist flow. As shown
in fig.\ \ref{FIG:3}c, $I_0$ decreases continuously with time and
drops down to a smaller and constant value once the sample flows.
Similar features are systematically measured, as shown in fig.\
\ref{FIG:4}, where the time evolution of $I_0$ for several
experiments is reported. In fig.\ \ref{FIG:4}, time is normalized
by $t_c$, and the absolute scattered intensity by $I$, the
intensity integrated over all angles in the flow regime
\cite{NoteIntensity}. We note that not only the time evolution but
also the absolute value of $I_0$ are very similar for all
experiments, which suggests that the evolution of $I_0$ reflects a
very general mechanism.  A peak of $I_0$ is nevertheless sometimes
observed, just before the collective reorientation of all
crystallites (at $t=t_c$) and well after the fluidization time. It
is presumably due to a transient orientation of some crystallites
(with tubes perpendicular to the flow) required to reach a more
favorable orientation as the stationary flow begins.

As pointed out above, the whole time evolution of the shear rate
(fig.\ \ref{FIG:1}) presents strong analogies with the strain rate
time evolution measured in the extension creep of composite
heterogeneous materials except that, in our case, the acceleration
of the rate (fluidization regime) is followed by a stationary flow
with low viscosity. By contrast, in extension creep of
heterogeneous materials, it is followed by the fracture of the
sample \cite{GuarinoEPJB2002, NechadPRL2005}. Based on this
analogy, we propose a phenomenological picture for our
experimental observations, which is inspired by the fiber bundle
models \cite{BFM}, and which explains the time evolution of both
$I_0$ and the shear rate, $\dot{\gamma}$. The initial state of our
material is powder-like, with a random and isotropic orientation
of the crystallites. We assume that "perpendicular crystallites",
i.e. crystallites for which the tubes are perpendicular to the
flow direction (those which contribute to $I_0$) form some sort of
percolated network, which resists flow. Upon application of a
stress, these crystallites progressively reorient, hence $I_0$
decreases, starting from the easiest ones (the reorientation being
more or less easy, based presumably on the local configuration
with the neighboring crystallites). Thus, we expect $\dot{\gamma}$
to decrease as the rate of reorientation decreases, until the
network formed by these crystallites is not percolated anymore.
This would sign the onset of the fluidization regime, where
$\dot{\gamma}$ increases and facilitates the reorientation of the
remaining perpendicular crystallites, hence $I_0$ continues to
decrease. Once all perpendicular crystallites are reoriented,
$I_0$ reaches a constant minimum value and the sample is expected
to flow with a constant $\dot{\gamma}$. To reinforce the analogy
between creep fracture and onset of flow, experiments with a
sufficient temporal and spatial resolution would be required in
order to determine the distribution of waiting times between
reorientation events and compare it with the power law
distributions measured in creep fracture
\cite{GuarinoEPJB2002,KunEPL2003}.

We thank L. Cipelletti, V. Trappe and L. Vanel for discussions, G.
Porte for a critical reading of the manuscript and E. Di Cola and
P. Panine for technical assistance during the SAXS experiments. T.
B. acknowledges partial support from the French-German
Collaborative Research Group 'Complex Fluids: from 2 to 3
Dimensions', jointly funded by the DFG (Germany), and the CEA and
CNRS (France). This work was supported in part by the NoE
``SoftComp`` (NMP3-CT-2004-502235) and by the ESRF.

%%%%%%%%%%%%%%%%%%%%%%%%%%%%%%%%%%


\begin{references}
\bibitem{NoteCrystallites} The standard Laue-Scherrer
relation relates the width of the Bragg peaks, $\Delta q$, to a
translational correlation length, $\xi$, which can be approximated
as the average size of the crystallites: $\xi \simeq 2 \pi /
\Delta q$. The full width at half maximum of the first order
diffraction peak is of the order of $8\times10^{-3}\,
\rm{nm^{-1}}$ (with a resolution of $3\times10^{-3}\,
\rm{nm^{-1}}$), hence $\xi \simeq 1.3 \, \mu
\rm{m}$.
\bibitem{NoteIntensity} Note that in the flow regime $I$ is largely dominated by the
intensity at $\theta = 90$ and $270$ deg .
\end{references}
\end{document}